%% file: main.tex
\documentclass{PoS}

\usepackage[caption=false]{subfig}
\usepackage{graphicx}
\usepackage{bm}
\usepackage{amssymb}
\usepackage{amsmath}
\usepackage{amsfonts}
\usepackage[utf8]{inputenc}
\usepackage{placeins}
\usepackage{booktabs}
\usepackage{gensymb}
\usepackage{microtype}
\usepackage{lineno}
\usepackage{multirow, makecell}

\usepackage{listings}
\usepackage{siunitx}
\usepackage{xspace}
\usepackage{floatrow}
\usepackage{comment}
\usepackage{resizegather}
\usepackage{textcomp}

\include{units}

\include{results}

\title{Tackling limited simulation and small signals}

\ShortTitle{Limited Simulation}

\author{Carlos A. Arg\"uelles\\
        Massachusetts Institute of Technology, Cambridge, MA 02139, USA\\
        E-mail: \email{caad@mit.edu}}
\author{\speaker{Austin Schneider}\\
        Dept. of Physics and Wisconsin IceCube Particle Astrophysics Center, University of Wisconsin, Madison WI 53706, USA\\
        E-mail: \email{aschneider@icecube.wisc.edu}}
\author{Tianlu Yuan\\
        Dept. of Physics and Wisconsin IceCube Particle Astrophysics Center, University of Wisconsin, Madison WI 53706, USA\\
        E-mail: \email{tyuan@icecube.wisc.edu}}
        
\abstract{We present a new, analytic, Poisson likelihood derived, technique to account for the statistical uncertainties inherent in simulation samples of limited size. This method has better coverage properties than other techniques, is valid for small data samples, and maintains good computational performance.}

\FullConference{36th International Cosmic Ray Conference -ICRC2019-\\
		July 24th - August 1st, 2019\\
		Madison, WI, U.S.A.}

\newcommand{\like}{\mathcal{L}}
\newcommand{\vectheta}{\vec{\theta}}
\newcommand{\vecw}{\vec{w}}
\newcommand{\prob}{\mathcal{P}}
\newcommand{\gprob}{\mathcal{G}}

\newcommand{\mcl}{\like_\textmd{Eff}}

\newcommand{\adhoc}{\mathcal{L}_{\textmd{AdHoc}}}

\newcommand{\agpar}{\alpha}
\newcommand{\bgpar}{\beta}

\newcommand{\meff}{m_\mathrm{Eff}}
\newcommand{\weff}{w_\mathrm{Eff}}

\begin{document}

For binned data, the Poisson likelihood is taken to be the probability to observe events in a bin and is commonly used in high-energy physics and particle astrophysics experiments.
Given an exact expectation rate, $\lambda(\vectheta)$, the probability of observing an integer $k$ events is
\begin{equation}
\label{eq:poisson}
\mathrm{
\like(\vectheta|k) = \mathrm{Poisson}(k;\lambda(\vectheta)) = \frac{\lambda(\vectheta)^{k}e^{-\lambda(\vectheta)}}{k!},}
\end{equation}
where $\vectheta$ is some set of physics parameters that determine $\lambda$. Given the stochastic nature of processes in particle physics, exactly determining $\lambda$ is often not possible and requires Monte Carlo (MC) simulations.
Simulation is often expensive, and so reweighting is employed as it enables a single simulation to be used to describe many physical hypotheses~\cite{Gainer:2014bta}. 

In such scenarios, an {\it ad hoc} likelihood is commonly used, where $\lambda$ is simply taken to be the sum of the weights in each bin, namely
\begin{equation} \label{eq:mcpoisson}
\adhoc(\vectheta|k) = \frac{\left(\sum_{i}{w_i(\vectheta)}\right)^{k}e^{-\left(\sum_{i}{w_i\left(\vectheta\right)}\right)}}{k!}.
\end{equation}
A notable downside of this {\it ad hoc} likelihood is that it neglects the statistical uncertainty inherent in estimating $\lambda(\vectheta)$ from a simulation of limited size.
For expensive simulations, or physical hypotheses far from the original simulation, this uncertainty can be non-negligible~\cite{Barlow:1993dm,Bohm:2013gla,Chirkin:2013lya,Glusenkamp:2017rlp}.
One can account for this uncertainty by incorporating a simulation derived prior on $\lambda$, denoted as $\prob\left(\lambda|\vecw(\vectheta)\right)$, that has non-zero variance. Thus, we can write the likelihood as the marginalization of the Poisson likelihood with the prior,
\begin{equation} \label{eq:generalpoisson}
\like_{\rm General}(\vectheta|k) = \int_{0}^{\infty}~\frac{\lambda^{k}e^{-\lambda}}{k!}\prob\left(\lambda|\vecw(\vectheta)\right)~d\lambda.
\end{equation}
We construct this prior based on the likelihood function of the simulation outcome and a prior on $\lambda$, $\prob(\lambda)$, 
\begin{equation} \label{eq:posterior}
\prob\left(\lambda|\vecw(\vectheta)\right) = \frac{\like(\lambda|\vecw(\vectheta))\prob(\lambda)}{\int_0^\infty \like(\lambda'|\vecw(\vectheta))\prob(\lambda')~d\lambda'},
\end{equation}
where in our implementation we have chosen we chosen $\prob(\lambda)$ to be uniform.

Let us first consider the case where all simulation events in the bin have equal weight.
In this scenario we can relate the number of events, $m$, and the weight of the events, $w$, to the quantities $\mu$ and $\sigma$, which are defined
\begin{equation}\label{eq:musigma}
\mu \equiv \sum_{i=1}^m w_i~\textmd{and}~\sigma^2 \equiv \sum_{i=1}^m w_i^2,
\end{equation}
and satisfy the relationships
\begin{equation}\label{eq:ids}
\mu=wm \textmd{,}~ \sigma^2=w^2 m \textmd{,}~ w=\sigma^2/\mu \textmd{, and }~ m = \mu^2/\sigma^2.
\end{equation}
The probability of obtaining $m$ events in the simulation bin can be modelled with the Poisson distribution assuming the true but unknown mean $\bar m$:
\begin{equation}\label{eq:mcprob}
\mathrm{Poisson}(M=m;\bar m) = \frac{e^{-\bar m} {\bar m}^m}{m!}.
\end{equation}
This allows us to rewrite the likelihood of $\lambda$ in terms of $\mu$ and $\sigma$ as
\begin{equation}
\like(\lambda|\vecw(\vectheta))=\like(\lambda|\mu, \sigma)=\frac{e^{-\lambda\mu/\sigma^2}\left(\lambda\mu/\sigma^2\right)^{\mu^2/\sigma^2}}{(\mu^2/\sigma^2)!}.
\label{eq:poisson_conditional}
\end{equation}

If the simulation event weights are not all equal, as is usually the case, then we can replace $w$ and $m$ with their ``effective'' counterparts $\weff$ and $\meff$.
These then relate to $\mu$ and $\sigma$ as
\begin{equation}\label{eq:effparameters}
\mu= \weff \meff~\textmd{and}~\sigma^2 = \weff^2 \meff.
\end{equation}
The replacement redefines the likelihood of our simulation outcome
\begin{align}
\label{eq:probmeff}
\like(\bar m|\meff)&= \frac{e^{-\bar m}{\bar m}^{\meff}}{\Gamma(\meff+1)},
\end{align}
which, assuming $\lambda = \weff {\bar m}$, can be rewritten as
\begin{equation}
\like(\lambda|\vecw(\vectheta))=\like(\lambda|\mu, \sigma)=\frac{e^{-\lambda\mu/\sigma^2}\left(\lambda\mu/\sigma^2\right)^{\mu^2/\sigma^2}}{\Gamma(\mu^2/\sigma^2+1)}.
\label{eq:poisson_conditional_arb}
\end{equation}
To simplify the notation, define
\begin{equation}\label{eq:alphabetamc}
\agpar \equiv \frac{\mu^2}{\sigma^2}+1~\textmd{and}~\bgpar \equiv \frac{\mu}{\sigma^2}.
\end{equation}
Substituting Eq.~\eqref{eq:poisson_conditional_arb} into Eq.~\eqref{eq:posterior} and assuming a uniform $\prob(\lambda)$, we obtain
\begin{align} \label{eq:theposterior}
\prob(\lambda|\vecw(\vectheta)) &= \bgpar \frac{ e^{-\lambda \bgpar}(\lambda \bgpar )^{\agpar-1}}{\Gamma(\agpar)}\nonumber \\
&= \frac{e^{-\lambda \bgpar } \lambda^{\agpar-1} \bgpar^{\agpar}}{\Gamma(\agpar)} \nonumber \\
&= \gprob(\lambda;\agpar, \bgpar),
\end{align}
where $\gprob(\lambda;\agpar, \bgpar)$ is the gamma distribution, with shape and inverse rate parameters $\alpha$ and $\beta$.
Finally, this can be substituted for $\prob\left(\lambda|\vec{w}(\vectheta)\right)$ in Eq.~(\ref{eq:generalpoisson}) so that
\begin{align}
\mcl(\vectheta|k) &=\int_{0}^{\infty}~\frac{\lambda^k e^{-\lambda}}{k!}\gprob(\lambda;\agpar, \bgpar)~d\lambda \\
&= \frac{\bgpar^\agpar\Gamma\left(k+\agpar\right)}{k!\left(1+\bgpar\right)^{k+\agpar}\Gamma\left(\agpar\right)} \\
&= \left(\frac{\mu}{\sigma^2}\right)^{\frac{\mu^2}{\sigma^2}+1}\Gamma\left(k+\frac{\mu^2}{\sigma^2}+1\right)\left[k!\left(1+\frac{\mu}{\sigma^2}\right)^{k+\frac{\mu^2}{\sigma^2}+1}\Gamma\left(\frac{\mu^2}{\sigma^2}+1\right)\right]^{-1}. \label{eq:parametrizedpoisson}
\end{align}

Equation \eqref{eq:parametrizedpoisson} is an effective likelihood, motivated by Poisson statistics, and derived with a Bayesian approach. It incorporates statistical uncertainties inherent in the MC approximation of the rate by encoding the distribution of weights in terms of $\mu$ and $\sigma^2$. The effective likelihood $\mcl$ can be easily substituted for $\adhoc$. A more thorough exposition, along with a generalization for different priors, $\prob(\lambda)$, is given in~\cite{Arguelles:2019izp}.

\bibliographystyle{ICRC}
\bibliography{references}

\end{document}

%% file: units.tex
\DeclareSIUnit \s {\second}
\DeclareSIUnit \MB {\mega\byte}
\DeclareSIUnit \GB {\giga\byte}
\DeclareSIUnit \TB {\tera\byte}
\DeclareSIUnit \PB {\peta\byte}
\DeclareSIUnit \Mbps {\mega\bit/\s}
\DeclareSIUnit \Gbps {\giga\bit/\s}
\DeclareSIUnit \Tbps {\tera\bit/\s}
\DeclareSIUnit \Pbps {\peta\bit/\s}
\DeclareSIUnit \kton {\kilo\tonne} 
\DeclareSIUnit \kt {\kilo\tonne}
\DeclareSIUnit \Mt {\mega\tonne}
\DeclareSIUnit \eV {\electronvolt}
\DeclareSIUnit \keV {\kilo\electronvolt}
\DeclareSIUnit \MeV {\mega\electronvolt}
\DeclareSIUnit \GeV {\giga\electronvolt}
\DeclareSIUnit \TeV {\tera\electronvolt}
\DeclareSIUnit \PeV {\peta\electronvolt}
\DeclareSIUnit \EeV {\exa\electronvolt}
\DeclareSIUnit \m {\meter}
\DeclareSIUnit \cm {\centi\meter}
\DeclareSIUnit \in {\inchcommand}
\DeclareSIUnit \km {\kilo\meter}
\DeclareSIUnit \kV {\kilo\volt}
\DeclareSIUnit \kW {\kilo\watt}
\DeclareSIUnit \MW {\mega\watt}
\DeclareSIUnit \MHz {\mega\hertz}
\DeclareSIUnit \mrad {\milli\radian}
\DeclareSIUnit \sr {\steradian}
\DeclareSIUnit \year {years}
\DeclareSIUnit \day {days}
\DeclareSIUnit \POT {POT}
\DeclareSIUnit \sig {$\sigma$}
\DeclareSIUnit\parsec{pc}
\DeclareSIUnit\lightyear{ly}
\DeclareSIUnit\foot{ft}
\DeclareSIUnit\ft{ft}
\DeclareSIUnit \ppb{ppb}
\DeclareSIUnit \ppt{ppt}
\DeclareSIUnit \samples{S}
\DeclareSIUnit \pe{PE}



%% file: results.tex
\usepackage{xparse}

\ExplSyntaxOn
\DeclareExpandableDocumentCommand{\eval}{m}{\fp_eval:n {#1}}
\ExplSyntaxOff

